\documentclass[12pt]{article}
\begin{document}
\title{Ultra High Energy Particles}
\author{B.G. Sidharth\\
International Institute for Applicable Mathematics \& Information Sciences\\
Hyderabad (India) \& Udine (Italy)\\
B.M. Birla Science Centre, Adarsh Nagar, Hyderabad - 500 063
(India)}
\date{}
\maketitle
\begin{abstract}
We revisit considerations of temporal order in relativistic effects,
taking into account Heisenberg's Uncertainty Principle. We then use
a formulation of relativistic Quantum Mechanical equations given by
Feshbach and Villars to exhibit novel particle antiparticle effects.
\end{abstract}
\section{Introduction}
In the context of the collision energies of a few $TeV$ being
attained at the LHC in CERN, Geneva we consider some ultra
relativistic effects, particularly for the Klein-Gordon (KG) and
Dirac equations. Following Weinberg \cite{weinbergrelcos} let us
suppose that in one reference frame $S$ an event at $x_2$ is
observed to occur later than one at $x_1$, that is, $x_2^0 > x_1^0$
with usual notation. A second observer $S'$ moving with relative
velocity $\vec{v}$ will see the events separated by a time
difference
$$x^{'0}_2 - x^{'0}_1 = \Lambda^0_\alpha (v) (x_2^\alpha -
x_1^\alpha)$$ where $\Lambda^\beta_\alpha (v)$ is the "boost"
defined by or,
$$x^{'0}_2 - x^{'0}_1 = \gamma (x_2^0 - x_1^0) + \gamma \vec{v} \cdot (x_2 - x_1)$$
and this will be negative if
\begin{equation}
v \cdot (x_2 - x_1) < - (x_2^0 - x_1^0)\label{2.13.1}
\end{equation}
We now quote from Weinberg \cite{weinbergrelcos}:\\
"At first sight this might seem to raise the danger of a logical
paradox. Suppose that the first observer sees a radioactive decay $A
\to B + C$ at $x_1$, followed at $x_2$ by absorption of particle
$B$, for example, $B + D \to E$. Does the second observer then see
$B$ absorbed at $x_2$ before it is emitted at $x_1$? The paradox
disappears if we note that the speed $|v|$ characterizing any
Lorentz transformation $\Lambda (v)$ must be less than unity, so
that (\ref{2.13.1}) can be satisfied only if
\begin{equation}
|x_2 - x_1| > |x_2^0 - x_1^0|\label{2.13.2}
\end{equation}
"However, this is impossible, because particle $B$ was assumed to
travel from $x_1$ to $x_2$, and (\ref{2.13.2}) would require its
speed to be greater than unity, that is, than the speed of light. To
put it another way, the temporal order of events at $x_1$ and $x_2$
is affected by Lorentz transformations only if $x_1 - x_2$ is
spacelike, that is,
$$\eta_{\alpha \beta} (x_1 - x_2)^\alpha (x_1 - x_2)^\beta > 0$$
whereas a particle can travel from $x_1$ to $x_2$ only if $x_1 -
x_2$ is timelike, that is,
$$\eta_{\alpha \beta}(x_1 - x_2)^\alpha (x_1 - x_2)^\beta < 0$$
"Although the relativity of temporal order raises no problems for
classical physics, it plays a profound role in quantum theories. The
uncertainty principle tells us that when we specify that a particle
is at position $x_1$ at time $t_1$, we cannot also define its
velocity precisely. In consequence there is a certain chance of a
particle getting from $x_1$ to $x_2$ even if $x_1 - x_2$ is
spacelike, that is, $|x_1 - x_2| > |x_1^0 - x_2^0|$. To be more
precise, the probability of a particle reaching $x_2$ if it starts
at $x_1$ is nonnegligible as long as
\begin{equation}
(x_1 - x_2)^2 - (x_1^0 - x_2^0)^2 \leq \frac{\hbar^2}{m^2}\label{X}
\end{equation}
where $\hbar$ is Planck's constant (divided by $2 \pi$) and $m$ is
the particle mass. (Such space-time intervals are very small even
for elementary particle masses; for instance, if $m$ is the mass of
a proton then $\hbar /m = w \times 10^{-14}cm$ or in time units $6
\times 10^{-25}sec$. Recall that in our units $1 sec = 3 \times
10^{10} cm$.) We are thus faced again with our paradox; if one
observer sees a particle emitted at $x_1$, and absorbed at $x_2$,
and if $(x_1 - x_2)^2 - (x_1^0 - x_2^0)^2$ is positive (but less
than $\hbar^2 /m^2$), then a second observer may see the particle
absorbed at $x_2$ at a time $t_2$ before the time $t_1$ it is
emitted at $x_1$".\\
To put it another way, the temporal order of causally connected
events cannot be inverted in classical physics, but in Quantum
Mechanics, the Heisenberg Uncertainty Principle leaves a loop hole.
To quote Weinberg again:\\
"There is only one known way out of this paradox. The second
observer must see a particle emitted at $x_2$ and absorbed at $x_1$.
But in general the particle seen by the second observer will then
necessarily be different from that seen by the first. For instance,
if the first observer sees a proton turn into a neutron and a
positive pi-meson at $x_1$ and then sees the pi-meson and some other
neutron turn into a proton at $x_2$, then the second observer must
see the neutron at $x_2$ turn into a proton and a particle of
negative charge, which is then absorbed by a proton at $x_1$ that
turns into a neutron. Since mass is a Lorentz invariant, the mass of
the negative particle seen by the second observer will be equal to
that of the positive pi-meson seen by the first observer. There is
such a particle, called a negative pi-meson, and it does indeed have
the same mass as the positive pi-meson. This reasoning leads us to
the conclusion that for every type of charged particle there is an
oppositely charged particle of equal mass, called its antiparticle.
Note that this conclusion does not obtain in nonrelativistic quantum
mechanics or in relativistic classical mechanics; it is only in
relativistic quantum mechanics that antiparticles are a necessity.
And it is the existence of antiparticles that leads to the
characteristic feature of relativistic quantum dynamics that given
enough energy we can create arbitrary numbers of particles and their
antiparticles".\\
As can be seen from the above, the two observers $S$ and $S'$ see
two different events, viz., one sees, in this example the protons
while the other sees neutrons. Moreover, this is a result stemming
from (\ref{X}), viz.,
\begin{equation}
0 < (x_1 - x_2)^2 - (x^0_1 - x^0_2)^2 (\leq
\frac{\hbar^2}{m^2})\label{Y}
\end{equation}
The inequality (\ref{Y}) points to a reversal of time instants
$(t_1,t_2)$ as noted above. However, as can be seen from (\ref{Y}), this happens
within the Compton wavelength.\\
We now consider the KG and Dirac equations in the above context of
time "reversals". It is well known that the KG relativistic equation
displays the phenomenon of negative energies. The problems of the KG
equation can be traced to the second time derivative. To avoid this
Dirac considers a first order equation, but here also there were
negative energies and he had to further propose his Hole Theory to
circumvent this, whereas Pauli and Wieskopf overcame the
difficulties by treating the KG equation in a field theoretical
sense, where the two degrees of freedom would represent distinctly
charged particles.
\section{The Feshbach Villars Formulation}
Feshbach and Villars \cite{feshbach} interpreted the KG equation in
a single particle rather than field theoretic context. Infact they
showed that this (F-V) formulation also applies to the Dirac
equation. To see this, we can rewrite the K-G equation in the
Schrodinger form, invoking a two component wave function,
\begin{equation}
\Psi = \left(\begin{array}{ll} \phi \\
\chi\end{array}\right),\label{2.16} \end{equation} The $K-G$
equation then can be written as (Cf.ref.\cite{feshbach} for details)
$$\imath \hbar (\partial \phi /\partial t) = (1/2m) (\hbar /\imath
\nabla - eA/c)^2 (\phi + \chi)$$
$$\quad \quad \quad +(e A_0 + mc^2)\phi,$$
\begin{equation}
\imath \hbar (\partial \chi / \partial t) = - (1/2m) (\hbar / \imath
\nabla - eA/c)^2 (\phi + \chi) + (e A_0 - mc^2)\chi\label{2.15}
\end{equation}
It will be seen that the components $\phi$ and $\chi$ are coupled in
(\ref{2.15}). In fact we can analyse this matter further,
considering free particle solutions for simplicity. We write, $$\Psi
= \left(\begin{array}{ll} \phi_0 (p) \\ \chi_0 (p)\end{array}\right)
\, e^{\imath / \hbar (p\cdot x-Et)}$$
\begin{equation}
\Psi = \Psi_0 (p)e^{\imath / \hbar (p\cdot x-Et)}\label{2.25}
\end{equation}
Introducing (\ref{2.25}) into (\ref{2.15}) we obtain, two possible
values for the energy $E$, viz.,
\begin{equation}
E = \pm E_p ; \quad E_p = [(cp)^2 +
(mc^2)^2]^{\frac{1}{2}}\label{2.26}
\end{equation}
The associated solutions are
$$\left.\begin{array}{ll} E = E_p  \quad \phi_0^{(+)} =
\frac{E_p+mc^2}{2(mc^2E_p)^{\frac{1}{2}}}\\
\psi_0^{(+)}(p): \quad \chi_0^{(+)} =
\frac{mc^2-E_p}{2(mc^2E_p)^{\frac{1}{2}}}\end{array}\right\}\phi_0^2
- \chi_0^2 = 1,$$
\begin{equation}
\left.\begin{array}{ll} E = - E_p  \quad \phi_0^{(-)} =
\frac{mc^2 - E_p}{2(mc^2E_p)^{\frac{1}{2}}}\\
\psi_0^{(-)}(p): \quad \chi_0^{(-)} = \frac{E_p +
mc^2}{2(mc^2E_p)^{\frac{1}{2}}}\end{array}\right\} \phi_0^2 -
\chi_0^2 = -1\label{2.27}
\end{equation}
It can be seen from this that even if we take the positive sign for
the energy in (\ref{2.26}), the $\phi$ and $\chi$ components get
interchanged with a sign change for the energy. Furthermore we can
easily show from this that in the non relativistic limit, the $\chi$
component is suppressed by order $(p / mc)^2$ compared to the $\phi$
component exactly as in the case of the Dirac equation \cite{bd}.
Let us investigate this circumstance further \cite{uheb,ness}.\\
It can be seen that (\ref{2.15}) are Schrodinger equations and so
solvable. However they are coupled. We have from them,
\begin{equation}
\dot{\phi} + \dot{\chi} = (eA_0 + mc^2) (\phi + \chi) - 2mc^2
\chi\label{wD}
\end{equation}
In the case if
\begin{equation}
mc^2 > > eA_0 \quad (\mbox{or} \, A_0 = 0)\label{wF}
\end{equation}
that is we are dealing with energies or interactions much greater
than the electromagnetic, (or in the absence of an external field)
we can easily verify that
\begin{equation}
\phi = e^{\imath px-Et} \mbox{and} \, \chi = e^{\imath
px+Et}\label{wG}
\end{equation}
is a solution.\\
That is $\phi$ and $\chi$ belong to opposite sign of $E (m \ne 0)$
(Cf. equation (\ref{2.27})). The above shows that the K-G equation
mixes the positive and negative energy solutions.\\
If on the other hand $m_0 \approx 0$, then (\ref{wD}) shows that
$\chi$ and $\phi$ are effectively uncoupled and are of same energy.
This shows that if $\phi$ and $\chi$ both have the same sign for
$E$, that is there is no mixing of positive and negative energy,
then the rest mass $m_0$ vanishes. A non vanishing rest mass
requires the mixing of both signs of energy. Indeed it is a well
known fact that for solutions which are localized, both signs of the
energy solutions are required to be superposed \cite{bd,schweber}.
This is because only positive energy solutions or only negative
energy solutions do not form a complete set.\\
Interestingly the same is true for localization about a time instant
$t_0$. That is physically, only the interval $(t_0 - \Delta t , t_0
+ \Delta t)$ is meaningful. This was noticed by Dirac himself when
he deduced his equation of the electron \cite{dirac}. Strictly speaking,
the electron would have the velocity of light, if we work with spacetime points.\\
In any case both the positive and negative energy solutions are
required to form a complete set and to describe a point particle at
$x_0$ in the delta function sense. The narrowest width of a wave
packet containing both positive and negative energy solutions, which
describes the spacetime development of a particle in the familiar
non-relativistic sense, as is well known is described by the Compton
wavelength. As long as the energy domain is such that the Compton
wavelength is negligible then our usual classical type description
is valid. In particular, the time inversion conditions stemming from
equation (\ref{X}) of Section 1 does not happen.\\
However as the energy approaches levels where the Compton wavelength
can no longer be neglected, then new effects involving
the negative energies and anti particles begin to appear (Cf.ref.\cite{feshbach}).\\
Further, we observe that from (\ref{wG})
\begin{equation}
t \to -t \Rightarrow E \to - E, \quad \phi \leftrightarrow
\chi\label{wJ}
\end{equation}
(Moreover in the charged case $e \to -e)$. It can be shown that the
Schrodinger equation goes over to the Klein-Gordon equation if we
allow $t$ to move forward and also backward in $(t_0 - \Delta t, t_0
+ \Delta t)$ (Cf.ref.\cite{ness}). Here we have done the reverse of
getting the Klein-Gordon equation into two Schrodinger
equations. This is expressed by (\ref{2.15}).\\
In any case we would like to reiterate that the two degrees of
freedom associated with the second time derivative can be
interpreted, following Pauli and Weisskopf as positive and
negatively charged particles or particles and anti particles.\\
\section{Remarks and Discussion}
We summarize the following:\\
i) From the above analysis it is clear that a localized particle
requires both signs of energy. At relatively low energies, the
positive energy solutions predominate and we have the usual
classical type particle behaviour. On the other hand at very high
energies it is the negative energy solutions that predominate as for
the negatively charged counterpart or the anti particles. More
quantitatively, well outside the Compton wavelength the former
behaviour holds. But as
we approach the Compton wavelength we have to deal with the new effects.\\
ii) To reiterate if we consider the positive and negative energy
solutions given by $\pm E_p$, as in (\ref{2.27}), then we saw that
for low energies, the positive solution $\phi_0$ predominates, while
the negative solution $\chi_0$ is $\sim (\frac{v}{c})^2$ compared to
the positive solution. On the other hand at very high energies the
negative solutions begin to play a role and in fact the situation is
reversed with $\phi_0$ being suppressed in comparison to $\chi_0$.
This can be seen from (\ref{2.27}).\\
iii) We could now express the foregoing in the following terms: It
is well known that we get meaningful probability currents and
subluminal classical type situations using positive energy solutions
alone as long as we are at energies low enough such that we are well
outside the Compton scale. As we near the Compton scale however, we
begin to encounter negative energy solutions or these
anti-particles.\\
From this point of view, we can mathematically dub the solutions
according to the sign of energy $(p_0/|p_0|)$ of these states: $+1$
and $-1$. This operator commutes with all observables and yet is not
a multiple of unity as would be required by Schur's lemma, as it has
two distinct eigen values. This is a superselection principle or a
superspin with two states and can be denoted by the Pauli matrices.
The two states would refer to the positive energy solutions and the
negative energy solutions (Cf.refs.\cite{uheb,ness}).\\
iv) We could now think along the lines of $SU (2)$ and consider the
transformation \cite{taylor}
\begin{equation}
\psi (x) \to exp [\frac{1}{2} \imath g \tau \cdot \omega (x)] \psi
(x).\label{4.2}
\end{equation}
This leads to a covariant derivative
\begin{equation}
D_\lambda \equiv \partial_\lambda - \frac{1}{2} \imath g \tau \cdot
\bar{W}_\lambda,\label{4.3a}
\end{equation}
as in the usual theory, remembering that $\omega$ in this theory is
infinitessimal. We are thus lead to vector Bosons $\bar{W}_\lambda$
and an interaction rather like the weak interaction. However we must
bear in mind that this new interaction between particle and
anti-particle \cite{report} would be valid only within the Compton
time, inside this Compton scale Quantum
Mechanical bridge.\\
v) We have already seen that even given the Lorentz transformation,
due to Quantum Mechanical effects, there could be an apparent
inversion of events, though at the expense of the exact description
of either observer. This has been brought out in Section 1 in the
case of the observer seeing protons and another seeing neutrons. We
now observe that in the above formulation for the wave function
$$\Psi = \left(\begin{array}{ll} \phi \\
\chi\end{array}\right),$$ $\phi$ (or more correctly $\phi_0$)
represents a particle while $\chi$ represents an antiparticle. So,
for one observer we have
\begin{equation}
\Psi \sim \left(\begin{array}{ll} \phi \\
0\end{array}\right)\label{B}
\end{equation}
and for another observer we can have
\begin{equation}
\Psi \sim \left(\begin{array}{ll} 0 \\
\chi\end{array}\right)\label{C}
\end{equation}
that is the two observers would see respectively a particle and an
antiparticle. This would be the same for a single observer, if for
example the particle's velocity got a boost so that (\ref{C}) rather
than (\ref{B}) would dominate after sometime.\\
Interestingly, just after the Big Bang, due to the high energy, we
would expect, first (\ref{C}) that is antiparticles to dominate,
then as the universe rapidly cools, particles and antiparticles
would be in the same or similar number as in the Standard Model, and
finally on further cooling (\ref{B}) that is particles or matter
would dominate.\\
vi) We now make two brief observations, relevant to the above
considerations. Latest results in proton-antiproton collisions at
Fermi Lab have thrown up the $Bs$ mesons which in turn have decayed
exhibiting CP violations in excess of the predictions of the
Standard Model, and moreover this seems to hint at a new rapidly
decaying particle. Furthermore, in these high energy collisions
particle to antiparticle and vice versa transformations have been
detected.

\end{document}